%% file: bmvc_maxwelling.tex
\documentclass{bmvc2k}

\title{Video Stream Retrieval of Unseen Queries using Semantic Memory}

\usepackage{amsmath,amsfonts,amssymb}
\usepackage{booktabs}
\usepackage{adjustbox}

\addauthor{Spencer Cappallo}{cappallo@uva.nl}{1}
\addauthor{Thomas Mensink}{tmensink@uva.nl}{1}
\addauthor{Cees G. M. Snoek}{cgmsnoek@uva.nl}{1}
\addinstitution{University of Amsterdam\\
Science Park 904\\
Amsterdam\\
The Netherlands}
\runninghead{Cappallo, Mensink, Snoek}{Video Stream Retrieval}

\def\eg{\emph{e.g}\bmvaOneDot}

\def\etal{\emph{et al}\bmvaOneDot}

\newcommand{\fig}[1]{Fig. \ref{fig:#1}}

\newcommand{\minisubsection}[1]{	\vspace{2mm}\noindent\textcolor{bmv@captioncolor}{\textbf{#1}\qquad}	}
\begin{document}

\maketitle

\begin{abstract}
Retrieval of live, user-broadcast video streams is an under-addressed and increasingly relevant challenge. The on-line nature of the problem requires temporal evaluation and the unforeseeable scope of potential queries motivates an approach which can accommodate arbitrary search queries. To account for the breadth of possible queries, we adopt a no-example approach to query retrieval, which uses a query's semantic relatedness to pre-trained concept classifiers. To adapt to shifting video content, we propose memory pooling and memory welling methods that favor recent information over long past content. We identify two stream retrieval tasks, instantaneous retrieval at any particular time and continuous retrieval over a prolonged duration, and propose means for evaluating them. Three large scale video datasets are adapted to the challenge of stream retrieval. We report results for our search methods on the new stream retrieval tasks, as well as demonstrate their efficacy in a traditional, non-streaming video task.
\end{abstract}

\input{intro-cs}
\input{relatedwork}
\input{streamretrieval}
\input{evaluation}
\input{experiments}

{\small
\bibliography{bmvc2016}
}
\end{document}

%% file: intro-cs.tex
\section{Introduction} \label{sec:intro}

This paper targets the challenge of searching among live streaming videos. This is a problem of increasing importance as more video content is streamed via services like Meerkat, Periscope, and Twitch. Despite the popularity of live streaming video, searching in its content with state-of-the-art video search methods, \eg \cite{trecvid,xu2015,jiang2015fast,mettes2016shuffle}, is nearly impossible as these typically assume the \emph{whole} video is available for analysis before retrieval. We propose a new method that can search across live video streams, for any query, without analyzing the entire video.

In live video, the future is unknowable thus one only has access to the past and present. 
It is therefore crucial to leverage knowledge of the (recent) past appropriately. 
Memory can be modeled with the aid of hidden Markov models or recurrent neural networks with long-short term memory. 
Through the ability to selectively remember and forget, recurrent neural networks have recently shown great potential for search in videos ~\eg~\cite{donahue2015longterm,ng2015beyond}. 
Inspired by the success of supervised memory models, we propose a mechanism to incorporate memory and forgetting in video stream retrieval without learning from examples.

Our search mechanism is founded on recent work in zero-shot classification, \eg \cite{norouzi2013zero,wu2014zero,jain2015objects}. 
These methods all use an external linguistic corpus to pre-train a semantic embedding such that terms which are used in similar contexts have similar vectors within the embedding~\cite{mikolov2013distributed}. 
Video frames are fed to a pre-trained convolutional neural network that outputs concept classification scores, which, after pooling over the entire video, can be projected into the embedding. 
An incoming text query utilizes the embedding to find the best matching videos. 
We adopt this established approach from the zero-shot community and re-purpose it for the new problem of live video stream retrieval.

\begin{figure}[t]
\centering
\includegraphics[width=.8\textwidth]{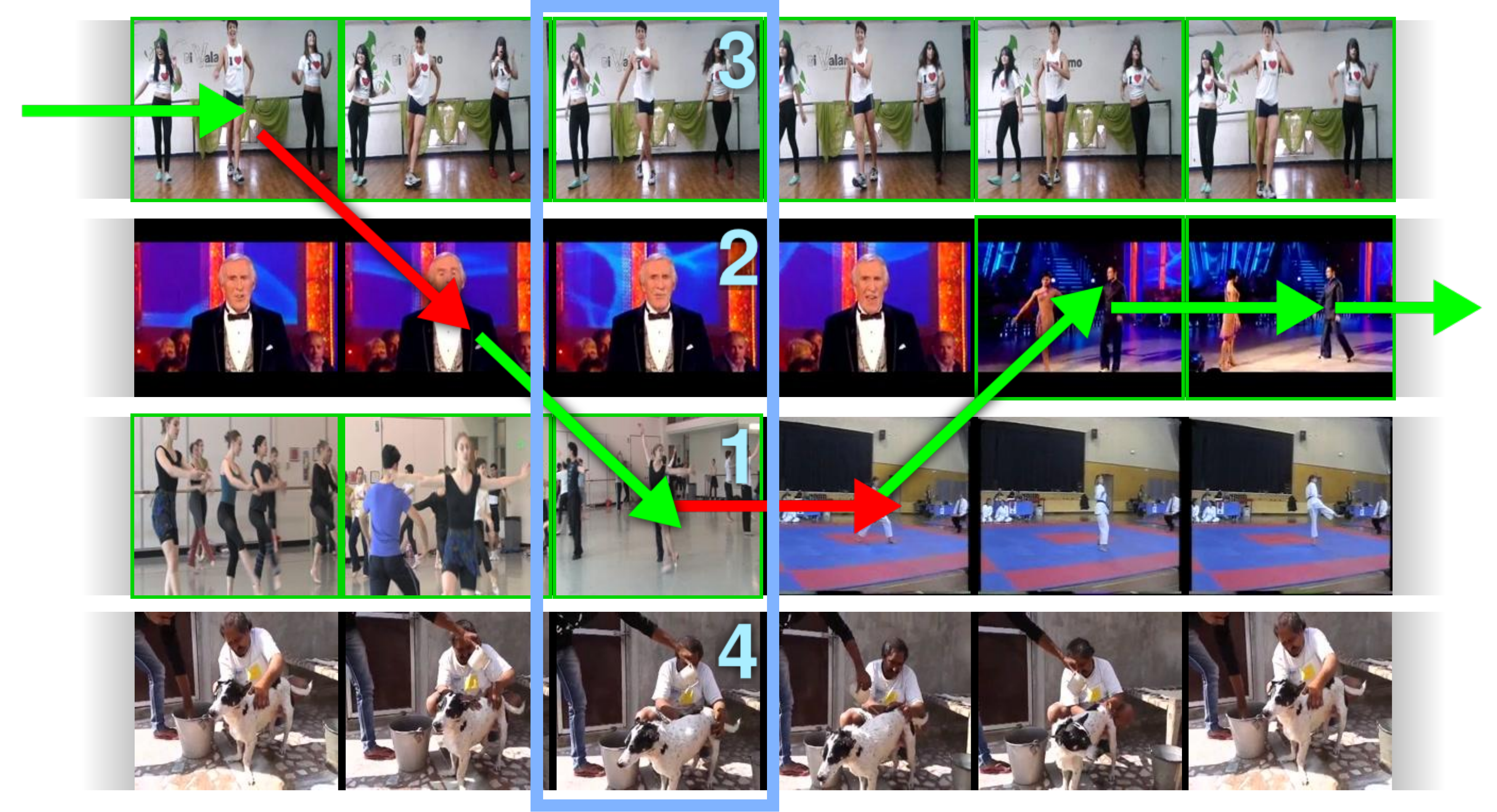}
\caption{Two video stream retrieval tasks: Instantaneous retrieval returns a ranked list of currently relevant streams, while continuous retrieval seeks to maximize the time spent on relevant streams, while minimizing the number of changes between streams.} 
\label{fig:tasks}
\end{figure}

We make three contributions. First, we establish the new problem of video stream retrieval and introduce a solution based on a framework popularized for zero-shot classification. Second, we introduce several methods to base retrieval on only the recent memory of the streams. Finally, in absence of any stream retrieval tasks in leading benchmarks such as ActivityNet~\cite{activitynet} and NIST TRECVID~\cite{trecvid}, we propose two evaluation settings  (see \fig{tasks}) based on publicly available datasets~\cite{activitynet,fcvid} and compare against established baselines. We also demonstrate that our method excels in more traditional whole-video retrieval scenarios. 

%% file: relatedwork.tex
\section{Related Work}
The setting of live stream retrieval is inherently related to a wide gamut of video tasks.
In this section we discuss some of the most relevant related work.

\minisubsection{Video Concept Detection} Video retrieval is aided by knowledge of the visual concepts which compose a scene and whose interaction through time define actions and events \cite{NATS05,smeaton2006,SnoekFNTIR09}.
Concept detection in video has been primarily addressed in the context of supervised classification tasks to detect objects, actions, and events, where the entire video is available for processing.
Most state-of-the-art approaches represent a video by pooling per-frame features extracted using a pre-trained convolutional neural network (CNN) ~\cite{jiang2010trecvid,xu2015,NagelBMVC15}. 
Such an approach is good for shorter, single-topic videos where the semantics of all frames are important for the final prediction, such as for events, actions, and activities \cite{trecvid,activitynet}. 
In a stream retrieval setting, however, the gestalt of the video in its totality becomes less important than what is currently happening in the stream.
Our approach instead emphasizes only recent information, by discarding past information which cannot be guaranteed to be relevant to the current stream content.

More explicit modelling of the temporal qualities of video has taken several forms, from temporal features, such as motion boundary histograms~\cite{wang2013dense}, to learning recurrent neural networks~\cite{baccouche2010action,donahue2015longterm,karpathy2014large,ng2015beyond}, and to localising the temporal extent of actions \cite{gaidon,JainCVPR14}. The idea of temporal windows have also been used to perform temporal action localisation in a hierarchical manner~\cite{oneata2013action}, which is orthogonal to any on-line stream processing concerns.

\minisubsection{Zero-shot prediction}
Live stream retrieval is a compelling use case for zero-shot prediction, given that the future content can not be predicted and accompanying descriptions can not be guaranteed.
Zero-shot classification seeks to transfer the models learned on one set of classes to another, related class through some intermediary knowledge source~\cite{lampert2009learning,jain2015objects,akata2013label,mensink2014costa,norouzi2013zero,mahajan2011joint}.
Most of these methods focus on a limited semantic transfer, \eg from a known set of animals to another set of animals; and trained on images and tested on images.
The yearly TRECVID MED zero-example benchmark~\cite{trecvid} has transferred this problem to event retrieval among videos, where a long, unseen textual event description is used to retrieve web videos. 
The scope of potential class types is also restricted in this case, and most participants use a fusion of aggregated event-related video features \cite{wu2014zero,chen2014event,Singh_2015_ICCV}.

One example of wider semantic transfer is the work of Jain \etal~\cite{jain2015objects}. 
Jain \etal exploit pre-trained ImageNet object detectors and an externally trained semantic embedding to transfer knowledge of ImageNet objects to actions and events in videos. 
This semantic embedding is constructed such that terms which are used in similar contexts have similar vectors within the embedding. Concept scores from a deep neural network trained to predict ImageNet classes are related to unseen concepts within the embedding space. Due to the encompassing nature of a broad linguistic corpus, this particular approach has been demonstrated to be useful for classifying a wide range of class abstractions, including objects \cite{norouzi2013zero}, actions, events \cite{jain2015objects}, and emoji \cite{emoji}. Such an approach is well-suited to the problem of live video stream retrieval, where possible queries may include these and many other class types.

%% file: streamretrieval.tex
\section{Video Stream Retrieval}
We focus on the novel problem of video stream retrieval. 
The nature of live, user-broadcast video has two major implications. 
First, the full range of potential future queries cannot be known, necessitating the ability to respond to unanticipated queries. 
Second, the future content of live video is unknown, and might not relate to prior content within the same stream, therefore we propose several methods to emphasize recent stream content. 



\subsection{Ranking Unanticipated Queries}

\begin{figure}[t]
\centering
\begin{minipage}[c]{.6\textwidth}
  \includegraphics[width=\textwidth]{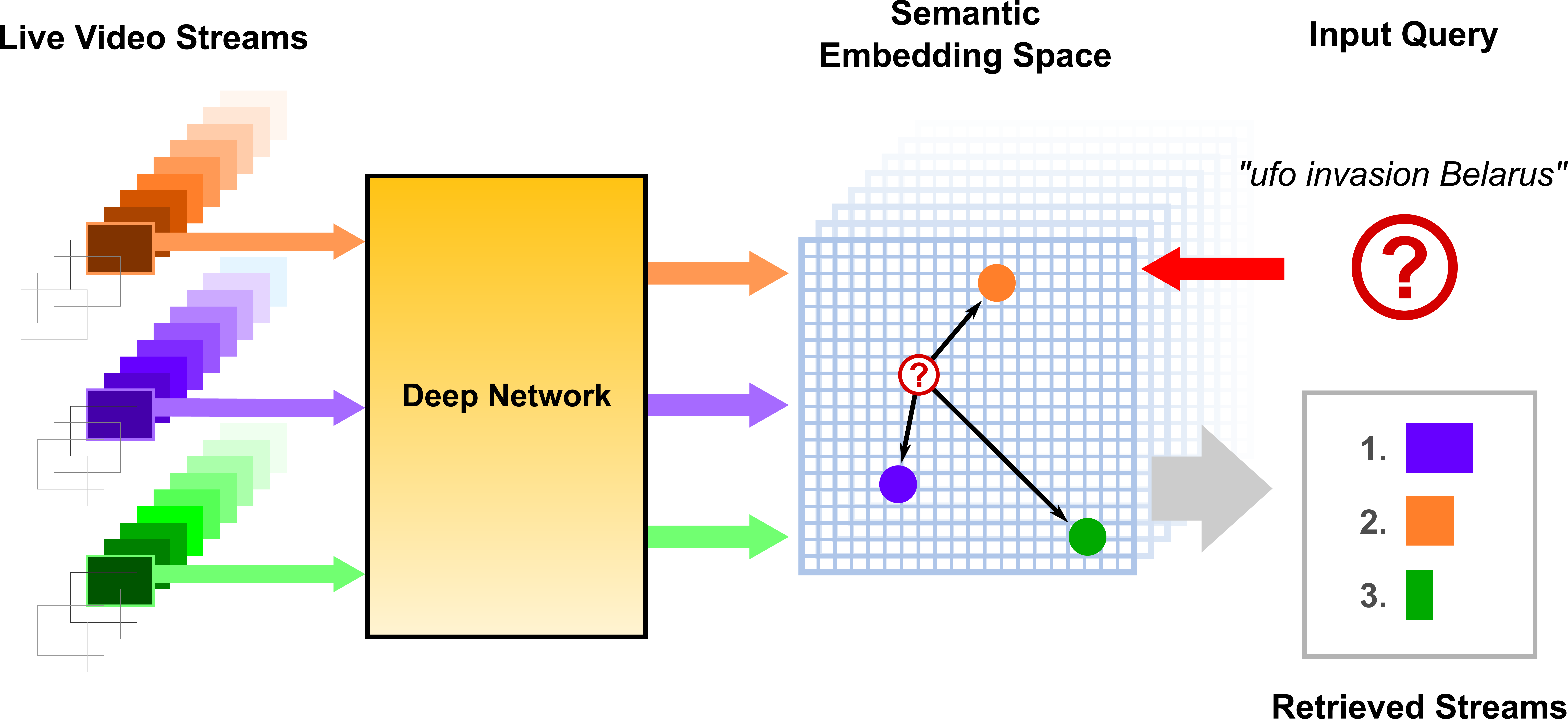}
\end{minipage}\quad
\begin{minipage}[c]{.34\textwidth}
  \caption{Stream retrieval for any query: Live streams are encoded by concept confidence scores, using a deep network. 
Streams are ranked based on the similarity between the query and these scores in a semantic space.}
\label{fig:streamretrieval}
\end{minipage}
\end{figure}

The goal is to retrieve relevant streams for a provided textual query $q$. 
To be robust against unanticipated queries, we follow a zero-shot classification paradigm~\cite{norouzi2013zero,jain2015objects,emoji}. 
A deep neural network trained to predict image classes is applied to the frames of the video stream as a feature extractor. 
$x_t$ represents the softmax output of the deep network across the output classes $C$ for a frame at time $t$. 
Some $\phi(x_t)$ encodes these concepts in a sparse manner. 
Both the concepts $C$ as well as the query $q$ are placed in a mutual embedding space (in our case, we use word2vec~\cite{mikolov2013distributed}), and video steams are scored based on the cosine similarities, using:
%
%
%
%
%
\begin{align}
\text{score}(q,x_t) = s(q)^\intercal \phi(x_t)
\label{eq:zeroshotretrieval}
\end{align}
where $s(q)$ returns a vector containing the cosine similarities between the embedding representation of the query $q$ and those of the concepts $C$. If the query $q$ comprises multiple terms, we use the mean of the per term scores, $s(q) = \frac{1}{N} \sum_{i=0}^N s(q_i)$, which has been shown to hold semantic relevance~\cite{mikolov2013distributed}. \fig{streamretrieval} shows the retrieval process.

\subsection{Memory for Stream Retrieval}
\label{memoryforstreamretrieval}

We introduce the notion of a "memory" for the problem of video stream retrieval, which aims to exploit recent information while limiting the effect of possibly irrelevant past information. 
The variable nature of live video means that past information might not be informative for future predictions, as a stream's content can change drastically. 
It is necessary to balance the utility of past information against the risk that it is no longer pertinent. 
In this section, we describe three approaches which use such a memory. 

\subsubsection{Memory Pooling}
Temporal pooling of frame-based features or concepts over an entire video is used in state-of-the-art approaches for standard video retrieval tasks~\cite{trecvid}. 
This strategy could be adapted to an on-line setting by pooling among all frames from time $t=0$ to the present. 
However, this introduces problems when the content of a stream changes, which is a particular concern with longer streams. 
For this reason, we pool instead over a fixed temporal memory $m$, which is tethered to the present and offers a restricted view on the past: 
\begin{align}
MP_{max}(x_t)   = \max_{i=t-m}^t x_i\qquad\qquad\qquad MP_{mean}(x_t)  = \frac{1}{m}\sum_{i=t-m}^t x_i
\end{align}
where $x_t$ denotes the features at time $t$, and we evaluate max pooling or mean pooling, denoted as $MP_{max}$ and $MP_{mean}$ respectively. 
At the start of a stream, when $m < t$, we instead use $m=t$. 
We set the memory duration $m$ through validation on a small set of queries which are disjoint from the test queries. 
The contribution of low confidence concepts introduces noisy predictions and influences the retrieval performance, therefore we use only the highest-valued pooled concepts, as proposed in~\cite{norouzi2013zero}. 

While mean and max pooling can be computed efficiently across all frames since $t=0$ in an iterative way, the introduction of a memory requires the storage of $m$ previous frames' worth of features for every concurrent stream. In part motivated by this expense, we introduce an alternative method which can be calculated in a stateless manner.

\subsubsection{Memory Welling}

\begin{figure}
\centering
\begin{minipage}[b]{.35\textwidth}
\centering
\raisebox{8mm}{
\includegraphics[width=0.8\textwidth]{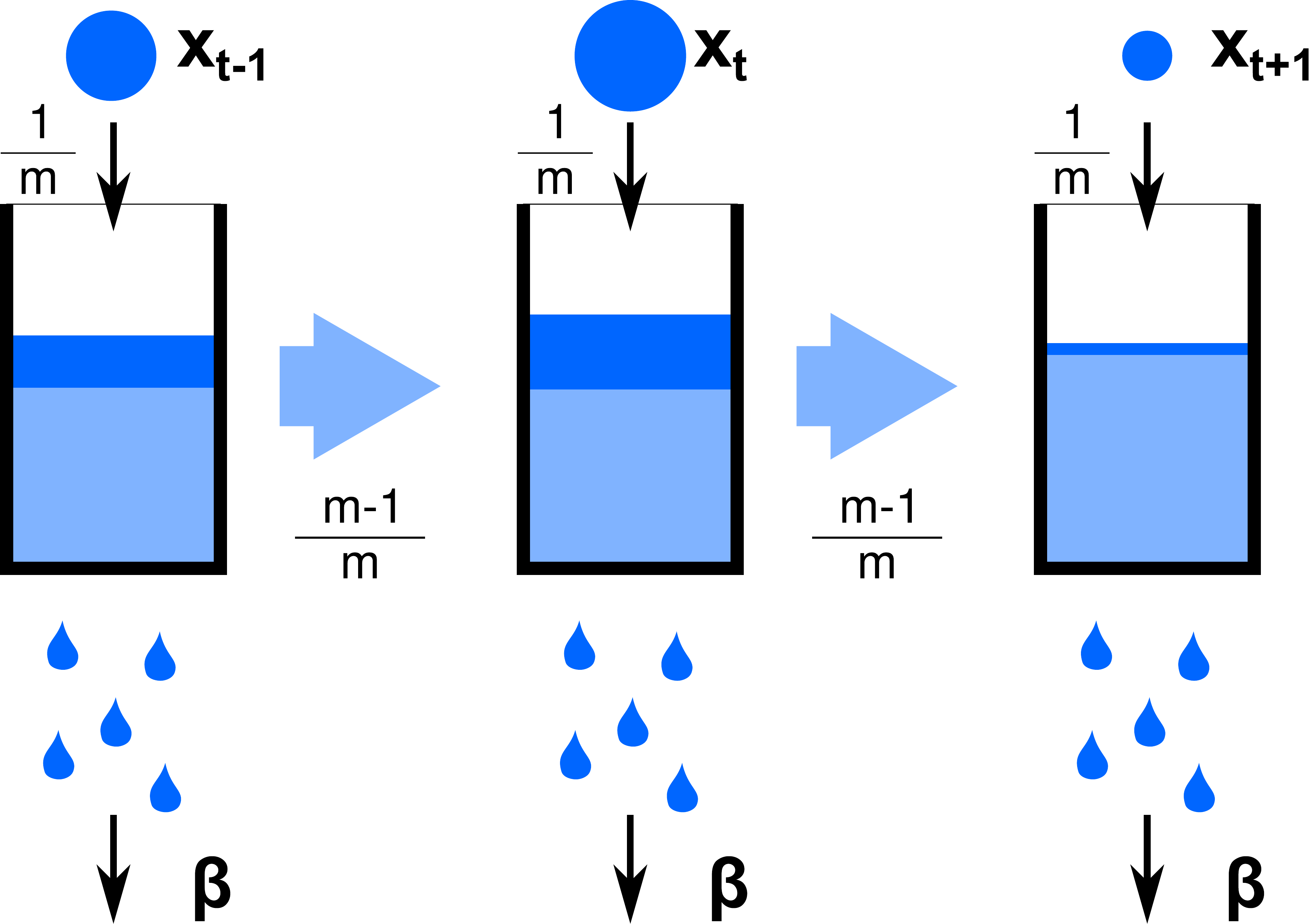}}\\
\end{minipage}
\begin{minipage}[b]{.64\textwidth}
\centering
\includegraphics[width=\textwidth]{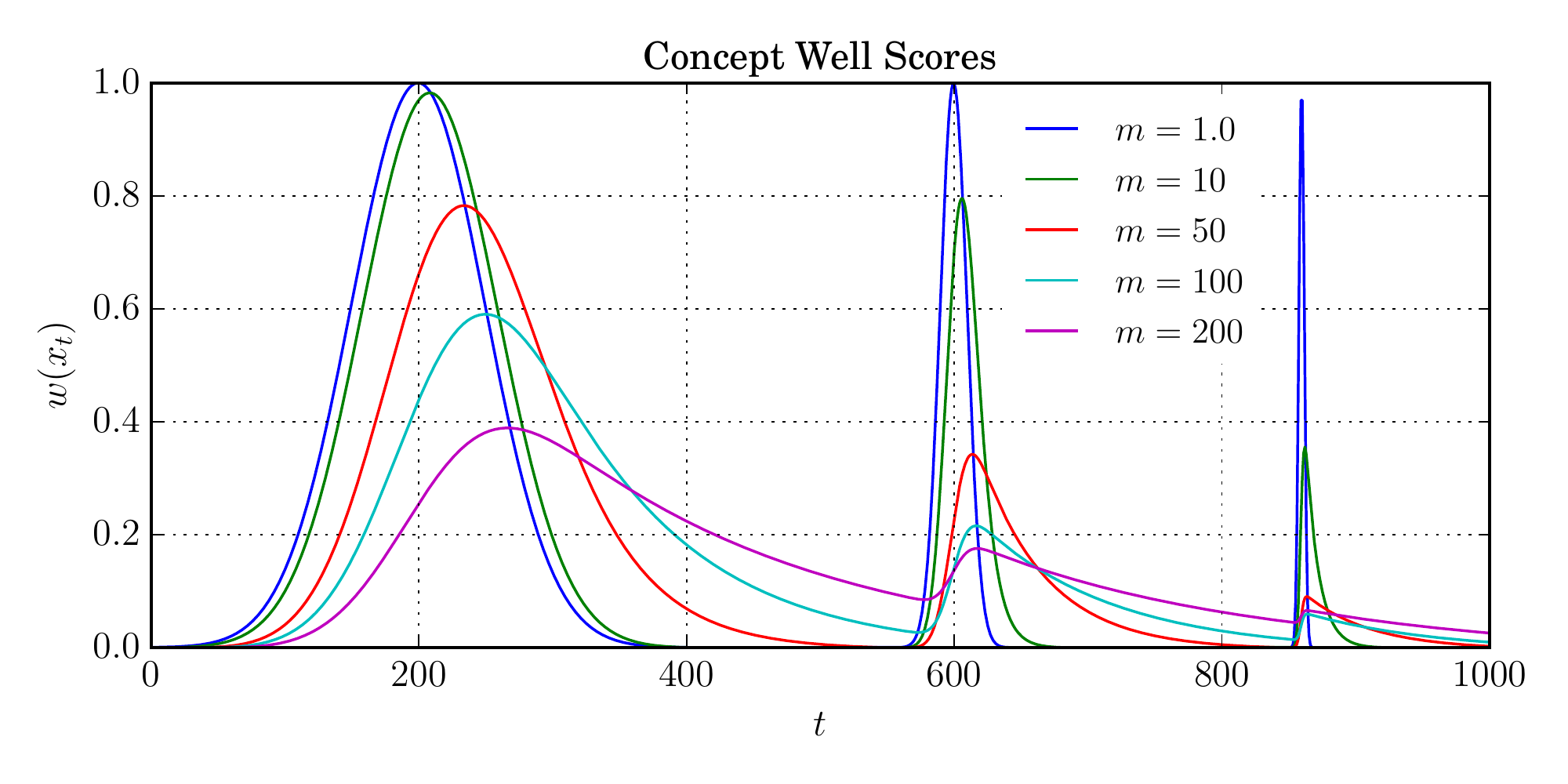}\\
\end{minipage}
\caption{ Left: Illustration of a memory well. New concept scores flow into the well, while old information gradually leaks out. Right: The effect of the memory parameter $m$ on memory welling. $m=1$ corresponds to the raw classifier confidence scores. Larger $m$ values result in a well which empties more slowly, but which is less responsive to sudden spikes.}
\label{fig:welling}
\end{figure}

The need to capture both long term trends and short-duration confidence spikes motivates the development of what we term \emph{memory wells}. 
In these wells, observations flow into the well at every timestep, but the well also leaks at every timestep, as illustrated in \fig{welling}. 
In contrast to the memory pooling, where all observations are weighed equally and observations beyond the memory horizon are lost, the impact of past observations on memory wells instead diminishes steadily over time. 
Memory wells are defined in the following manner:
\begin{align}
w(x_t) = \max\left(\frac{m-1}{m} w(x_{t-1}) + \frac{1}{m} x_t - \beta, 0\right),
\end{align}
where the current value of a well relies on the well's value at time $t-1$, diminished by a tunable memory parameter $m$ and a fixed constant leaking term $\beta$. 
We illustrate the effect of $m$ in \fig{welling}. Note that $m$ in this formulation is somewhat different from that in the memory pooling approach, albeit both aim to tune the contribution of past frames. Memory wells bear a faint resemblance to stacks or queues, but are distinguished by being unordered aggregrations of continuous values rather than ordered collections of discrete items, and by their discarding of stale data over time through leakiness.

The $\beta$ term creates sparseness in the representation, which ensures that only recent or consistently present concepts are used for prediction. 
We fix $\beta = \frac{1}{C}$, where $C$ is the number of concepts, as this is a lower bound for a concept being present at time $t$. This is the value that the classifier would output for every concept if it considered them all equally likely to be present in the current frame. Enforcing sparseness, or rather, enforcing reliability of concept scores, means that the memory well values can be used directly in Equation~\ref{eq:zeroshotretrieval}, without the need to arbitrarily select some number of the highest-confidence concepts. 

\minisubsection{Max Memory Welling}
In the case of short streams and traditional video processing tasks, which are likely to have more consistent content, the short-term nature of memory welling can be a limitation, even if its properties are still effective for improving temporally local predictions. 
Memory welling can be adapted to this task through temporal max pooling across the query scores per stream:
\begin{align}
\text{score}(q,x_t) = \max_{i=0}^t \left(s(q)^\intercal w(x_i)\right)
\end{align}
%
%
This exploits temporally local, high confidence predictions from the welling approach, which might be averaged away in traditional whole video pooling. It is well-suited to single-topic content such as short streams and traditional, full video retrieval tasks.

\minisubsection{Computational Complexity} The proposed approach for stream retrieval, particularly with memory welling, is comparatively lightweight, which is important for the targeted setting. Semantic similarity values, $s(\cdot)$, can be pre-computed and hashed for the entire query vocabulary, therefore $s(q)$ scales at $\mathcal{O}(l)$, where $l$ is the number of terms in the query $q$. Calculating $s(\cdot)^\intercal w(x_t)$ depends on the number of concepts, therefore has a complexity of $\mathcal{O}(m)$ for one stream, where $m$ is the number of concepts. Across $n$ streams, this gives a total complexity of $\mathcal{O}(lmn)$. $x_t$ has a constant cost per frame, which on a modern GPU is below $80$ms per batch of 128 frames. As $l$ and $m$ are fixed and relatively small constants, real time stream retrieval with the proposed method and a reasonable sampling rate is achievable.

%% file: evaluation.tex
\section{Tasks for Video Stream Retrieval}
To reflect the on-line nature and the diverse applications of video stream retrieval we propose two evaluation settings: 
\emph{i}) Instantaneous Retrieval, which measures the retrieval performance at any given time $t$; and 
\emph{ii}) Continuous Retrieval, where a succession of streams relevant to a single query are retrieved over a prolonged duration.

\subsection{Instantaneous Retrieval}

The goal of instantaneous retrieval is to retrieve the most relevant stream for a query $q$ at any arbitrary time $t$. 
This temporal assessment is important, given that a model which only performs well when a stream has ended is useless for discovery of live video streams.

To incorporate the temporal domain, we use the mean of the average precision (AP) scores per time step $t$, which we coin Temporal Average Precision (TAP). 
Letting $\text{AP}_t$ denote the AP score for some query at time $t$, the TAP then corresponds to the mean $\text{AP}_t$ across all times for which there is at least one relevant stream:
\begin{align}
\text{TAP} = \frac{1}{\sum_t y^t}\sum_t \text{AP}_t\cdot y^t,
\end{align}
where $y^t$ indicates whether there is at least one relevant stream for the query at time $t$.

\subsection{Continuous Retrieval} 
The goal of the continuous retrieval task is to maximize the fraction of time spent watching relevant streams, while minimizing the number of times the stream is changed. 
Consider a viewer searching for coverage of the Olympics. When one stream stops showing the Olympics, she wants to switch to another stream showing the Olympics. However, switching between two streams every second, even if both relevant, provides a poor viewing experience.

To evaluate this scenario, we consider the number of \emph{zaps}. 
A zap is any change in the retrieved stream or its relevancy, including the move at time $t=0$ to the first retrieved stream. 
We distinguish good zaps, which is any zap that moves from a currently irrelevant stream to a currently relevant stream, from all other (bad) zaps.
The count of good zaps and bad zaps are represented by $z_+$ and $z_-$. 

The fraction of good zaps to total zaps, $\frac{z_+}{z_+ + z_-}$, describes the average quality of individual changes, but offers an incomplete picture of the system's temporal consistency. 
Imagine a system which only ever retrieves one stream, which is initially relevant but quickly turns and remains irrelevant. Despite its performance, this would achieve a score of 0.5, as it would have had one good zap over a total of two zaps. 
To incorporate overall accuracy over time, we also reward an algorithm choosing to correctly remain on a relevant stream. 
Letting $r_+$ track the number of times an algorithm remains on relevant stream, the \emph{zap precision} ZP is
\begin{align}
\text{ZP} = \frac{z_+ + r_+}{\sum_t y^t}
\end{align}
%
%
where $y^t$ again represents whether or not there is at least one relevant stream at time $t$. 

%% file: experiments.tex
\section{Experiments}
\subsection{Setup}
\noindent\textcolor{bmv@captioncolor}{\textbf{Datasets}\qquad}
We evaluate our methods on three large scale video datasets: 
i) ActivityNet~\cite{activitynet} (AN), a large action recognition dataset with 100 classes and 7200 labeled videos. Performance is evaluated on a test set composed of 60 classes randomly selected from the combined ActivityNet training and validation splits, and a validation set of the other 40 classes is used for parameter search;
ii) A subset of the Fudan-Columbia Videos~\cite{fcvid} (coined FCVS), composed of 25 videos for each of the 239 classes making up 250 hours of video, which we split into a validation set of 50 classes and a test set of 179 classes. FCVS annotations are more diverse (objects, locations, scenes, and actions), but lack temporal extent, so a class is assumed to be relevant for the duration of a video;
iii) TRECVID MED 2013~\cite{trecvid} (MED), an event recognition dataset, used to evaluate the efficacy of our memory-based approach against published results. To facilitate comparison, the setting used by \cite{jain2015objects} is replicated: whole-video retrieval using only the event name.

In addition to evaluating on short web videos themselves, we introduce AN-L and FCVS-L, which are adaptations to simulate longer streams with varied content. To accomplish this, individual videos are randomly concatenated until the simulated stream is at least 30 minutes long. Annotations from the original videos are propagated to these concatenated videos.
Details of the data set splits will be made available to allow future comparison\footnote{http://staff.science.uva.nl/s.h.cappallo/data.html}. 

\minisubsection{Features}
We sample videos at a rate of two frames per second. Each frame is represented by the softmax confidence scores of 13k ImageNet classes, which are extracted using a pre-trained deep neural network from~\cite{mettes2016shuffle}. 
The network was trained on ImageNet~\cite{imagenet} and its structure is based on the GoogLeNet network~\cite{szegedy2015going}. 
Our semantic embedding is a 500-dimensional skip-gram word2vec \cite{mikolov2013distributed} model trained on the text accompanying 100M Flickr images \cite{yfcc100m}, similar to the one used in~\cite{emoji,jain2015objects}.

\minisubsection{Evaluation and Baselines}
To simulate the streaming setting, performance is evaluated sequentially across all videos, using only the present and past frames. 
Results are reported in the previously described TAP and ZP metrics averaged over all test classes. 
For the memory based methods, the optimal value of $m=m^{*}$ is determined on the validation set containing videos of classes not present in the test set. 
The two extremes of memory pooling are used as baselines: $m=1$, which simply relies on the current frame of a video to make a prediction; and $m=t$, which corresponds to pooling over the entirety of the stream up to the present time, similar to whole-video pooling used in video retrieval scenarios \cite{trecvid}. This approach has also been explored as a basis for whole video action recognition \cite{fernando2015modeling}.

\subsection{Instantaneous and Continuous Stream Retrieval}
\input{experiments_table}

We first compare our proposed methods and baselines on the instantaneous and continuous stream retrieval tasks. 
Table \ref{tab:results_overview} shows the results for the two tasks. 
In general, we observe that memory-based approaches shine when query relevance is temporally limited, as in the AN, AN-L, and FCVS-L datasets. For a setting like FCVS, where a single annotation covers an entire stream, the baselines become more competitive. In a scenario where streams are guaranteed to be short in duration and focus on a single topic, then a max memory welling approach makes the most sense. For streams of indeterminate length and content, the memory welling approach offers the best results and flexibility to cover any situations that may arise. 
In the continuous retrieval setting, the $m=t$ baselines and max memory welling perform poorly, likely due to their inability to respond quickly to changes in stream content. 

\begin{figure}
\begin{minipage}{0.5\textwidth}
\centering
\includegraphics[width=\textwidth]{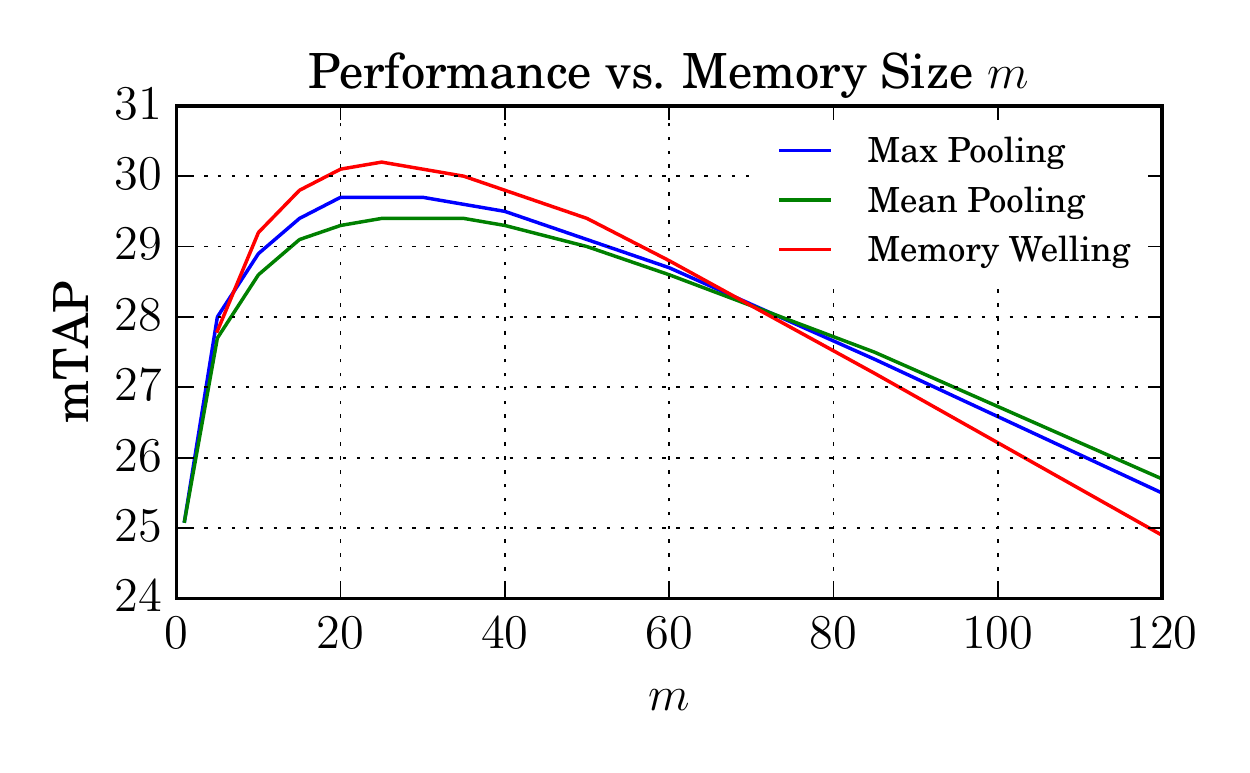}
\end{minipage}
\begin{minipage}{0.48\textwidth}
\centering
\caption{Effect of $m$ parameter on the ActivityNet-Long (AN-L) dataset. All approaches share a similar dependence, with a peak around $m=25$, which corresponds to 12.5 seconds at our sampling rate. Past this point, the irrelevancy of past information becomes overpowering.}
\label{fig:mperformance}
\end{minipage}
\end{figure}

\minisubsection{Impact of Memory Length}
The impact of the $m$ parameter on instantaneous retrieval is shown in Figure \ref{fig:mperformance}. The response of memory-based approaches to changing content degrades if $m$ is too large, and its resistance to noisy spikes suffers if $m$ is too small. $m$ values between 15 and 35 appear to be most adequate for the identification of current content. 

%

\begin{table}
\centering
\begin{minipage}[c]{0.43\textwidth}
\footnotesize
\begin{tabular}{l r r}
\toprule
\bfseries Category 	&\bfseries MMW 	& \bfseries MMP\\\midrule
Art 				& 20.4					& 14.7\\
Leisure \& Tricks 	& 34.0					& 24.5\\ 
Nature 				& 64.6					& 55.2\\
Travel 				& 31.3					& 30.0\\
Everyday Life 		& 31.2					& 21.0\\
Sports 				& 48.5					& 32.6\\
Beauty \& Fashion 	& 24.3					& 17.1\\
Music 				& 35.7 					& 28.8\\
DIY 				& 16.9					& 13.1\\
Education \& Tech 	& 67.8					& 51.4\\
Cooking \& Health 	& 27.7					& 20.9\\
\toprule
\bfseries Annotation Type 				&\bfseries MMW 	& \bfseries MMP\\\midrule
Place - {\tiny Particular location }	& 39.1					& 26.8\\
Object - {\tiny Thing or creature }		& 67.1					& 50.0\\
Scene - {\tiny Generic scene setting}	& 39.4					& 33.3\\
Event - {\tiny Particular occurrence}	& 28.5					& 21.1\\
Activity - {\tiny Human activities}		& 30.3					& 22.2\\
\bottomrule
\end{tabular}
\end{minipage}
\begin{minipage}[c]{0.55\textwidth}
\centering
\caption{Instantaneous retrieval on FCVS by annotation category and type for Max Memory Welling and Max Memory Pooling. The query type significantly affects the retrieval quality, but welling yields improvement in all cases. Below: Per-class scatterplot comparison of the results.}
\label{fig:categoryperformance}
\includegraphics[width=\textwidth]{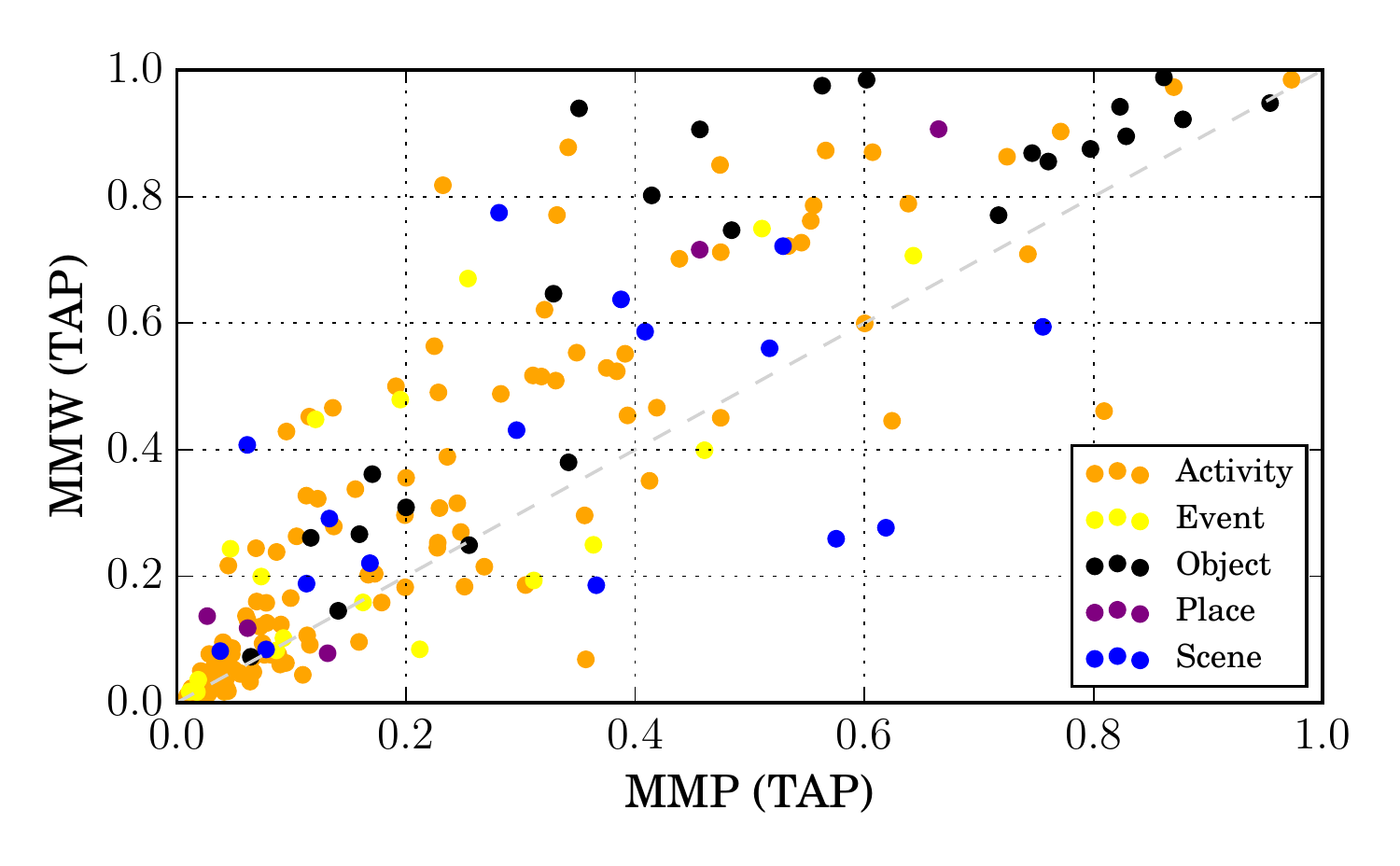}
\end{minipage}

\end{table}

\minisubsection{Per-Category Performance}
For the FCVS dataset, we report the performance per category and per annotation type in Table \ref{fig:categoryperformance}. 
The categories are provided within the annotation hierarchy, while we have manually assigned the FCVS test classes to one of five types. 
The Nature, Education \& Tech, and Sports categories perform strongly, likely due to their domain similarity with the ImageNet concepts used to train the deep network. This is also illustrated by the strong performance of the Object type classes. Meanwhile, the Art and DIY categories perform very poorly. The videos within these categories depict many hard-to-distinguish activities. For example, DIY contains four different classes which are composed primarily of video of hands manipulating paper. This very similar visual content is challenging. Furthermore, Events and Activities prove difficult to retrieve, likely due to their reliance on time. 
This highlights the difficulty of representing queries with an intrinsic temporal element through constituent static image concepts (such as ImageNet concepts).

\begin{table}[t]
\begin{center}
\begin{minipage}{0.48\textwidth}
\footnotesize
\begin{tabular}{l r r}
\toprule
\bfseries Method &\bfseries  mAP (\%) &\bfseries mTAP\\\midrule
Chen \etal \cite{chen2014event} & 2.4\\
Wu \etal \cite{wu2014zero} & 3.5\\
Jain \etal \cite{jain2015objects} & 3.5 \\\midrule
Jain \etal (our features) & 3.5 & 9.2\\
Max Memory Welling & \bfseries 4.7 & \bfseries 17.8\\\bottomrule
\end{tabular}
\end{minipage}
\begin{minipage}{0.5\textwidth}
\centering
\caption{Performance of Max Memory Welling on the TRECVID2013 MED task. MMW outperforms the state-of-the-art on this task. As \cite{jain2015objects} uses a different deep network, we also verify their results with our features. Further, we compare the performance of such an approach on instantaneous retrieval.}
\label{table:trecvid}
\end{minipage}
\end{center}
\end{table}

\subsection{Whole Video Retrieval}
To compare our method against published results, we report mAP results on the MED dataset, following the setting from~\cite{jain2015objects}: multimedia event retrieval based solely on the event-name. 
The results are shown in Table \ref{table:trecvid}, where we also compare to the visual-only results of \cite{chen2014event,wu2014zero}.
Our Max Memory Welling outperforms these methods, while being on par with the more advanced Fisher Vector event-name encoding (4.2 \% mAP) of~\cite{jain2015objects}. 
Note, such an event-name encoding could also be used alongside our method. 
The Max Memory Welling approach is able to leverage short-term, high-confidence predictions generated through memory welling, which is useful for whole video retrieval. 

\section{Conclusion}
The retrieval of live video streams requires approaches which can respond to unanticipated queries. We present such an approach, and demonstrate the importance and utility of memory-based methods, such as memory welling, for both on-line stream retrieval and other zero-example video tasks. We explore two scenarios, instantaneous and continuous retrieval, that follow naturally from the problem of stream retrieval, and offer an approach for evaluating these scenarios on existing, abundant large scale video datasets. 

\vspace{1em}
{\small\noindent\textcolor{bmv@captioncolor}{\textbf{Acknowledgements}}\newline%
This research is supported by the STW STORY project and the NWO VENI What\&Where project.}

%% file: experiments_table.tex
\begin{table}[t]
\centering
\caption{Results of instantaneous and continuous retrieval across all datasets and tasks. $m=1$ corresponds to using only the current frame, while $m=t$ means that pooling is performed over all past and present frames. Memory welling offers the best performance flexibility.}
\label{tab:overview}
\label{tab:results_overview}\vspace{0.2em}
{\small
\begin{tabular*}{\textwidth}{l @{\extracolsep{\fill}} r r r r r r }
\toprule
 & \multicolumn{4}{c}{\textcolor{bmv@captioncolor}{\bfseries Instantaneous (\% TAP)}}&\multicolumn{2}{c}{\textcolor{bmv@captioncolor}{\bfseries Continuous (\% ZP)}} \\\cmidrule(r){2-5}\cmidrule(r){6-7}
 				&\bfseries AN 	&\bfseries FCVS &\bfseries AN-L &\bfseries FCVS-L 	&\bfseries AN-L &\bfseries FCVS-L \\\cmidrule(r){1-5}\cmidrule(r){6-7}
Random 			& 1.4			&4.9			&	3.6			&	2.9				& 1.3			&	1.1		\\ \cmidrule(r){1-5}\cmidrule(r){6-7}
Mean Memory Pooling \\
$\quad m=1$		& 16.9			& 21.4			& 25.1			& 24.8				& 21.9			& 21.6\\
$\quad m=t$		& 18.4			& 30.7			& 8.5			&	9.3				& 5.9			& 6.3 \\
$\quad m=m^*$		& 21.7			& 28.8			& 29.3			& 30.0				& 27.5			& 27.7\\ \cmidrule(r){1-5}\cmidrule(r){6-7}
Max Memory Pooling \\
$\quad m=t$		&  20.0			& 27.4			& 9.0			& 9.5				& 5.9			& 6.0\\
$\quad m=m^*$	& 21.0			& 27.5			& 29.7			& 30.3				& 27.3			& 27.5 \\ \cmidrule(r){1-5}\cmidrule(r){6-7}
Memory Welling  & 22.5			& 30.5			&\bfseries 30.1	&\bfseries 30.6		&\bfseries 28.3	&\bfseries 28.4 \\
Max Memory Welling &\bfseries 24.6 &\bfseries 35.9 & 11.0		& 	15.9				& 5.6			&10.9 	\\	 
\bottomrule
\end{tabular*}
}
\end{table}